\documentclass[fleqn,10pt]{wlscirep}
\usepackage[utf8]{inputenc}
\usepackage[T1]{fontenc}

\usepackage{graphicx}
\usepackage{physics}

\title{Transitions between phyllotactic lattice states in curved geometries}

\author[1]{H. S. Ansell}
\author[2]{A. A. Tomlinson}
\author[2,*]{N. K. Wilkin}

\vspace{-1em}
\affil[1]{Department of Physics and Astronomy, University of Pennsylvania, Philadelphia, PA, 19104, USA}
\affil[2]{School of Physics and Astronomy, University of Birmingham, Birmingham, B15 2TT, UK}
\affil[*]{n.k.wilkin@bham.ac.uk}
\date{\vspace{-1em}\normalsize \today}

\begin{abstract}
Phyllotaxis, the regular arrangement of leaves or other lateral organs in plants including pineapples, sunflowers and some cacti, has attracted scientific interest for centuries. More recently there has been interest in phyllotaxis within physical systems, especially for cylindrical geometry. In this letter, we expand from a cylindrical geometry and investigate transitions between phyllotactic states of soft vortex matter confined to a conical frustum. We show that the ground states of this system are consistent with previous results for cylindrical confinement and discuss the resulting defect structures at the transitions. We then eliminate these defects from the system by introducing a density gradient to create a configuration in a single state. The nature of the density gradient limits this approach to a small parameter range on the conical system.  We therefore seek a new surface, the horn, for which a defect-free state can be maintained for a larger range of parameters.
\end{abstract}

\begin{document}
\flushbottom
\maketitle

\section{Introduction}
Novel defect structures can form in systems for which the lattice structure is not commensurate with the confining geometry. A regular triangular lattice on a sphere requires twelve positive disclinations, while toroidal crystals exhibit arrangements with negative and positive disclinations on the interior and exterior of the torus respectively~\cite{Giomi2008a, Giomi2008b}. Moreover, close packed particles on curved surfaces, at large enough system size, exhibit additional defects in the form of grain boundary scars~\cite{Bausch2003, Lipowsky2005, Sausset2010, Einert2005, Burke2015, Soni2018} or pleats~\cite{Irvine2010} in order to relieve the lattice strain energy.

Under cylindrical confinement, objects including discs~\cite{Mughal2014}, spheres~\cite{Erickson1973, Pickett2000, Lohr2010, Mughal2012, Fu2016}, nanoparticles~\cite{Kholbystov2004, Yamazaki2008, Liang2014} and radially repulsive point-like particles~\cite{Oguz2011, Tomlinson2020} arrange into regular structures which can be described as phyllotactic states. Phyllotaxis, most commonly used to describe the regular arrangement of leaves or other lateral organs in plants, describes the arrangement of objects into regular spirals which intersect to form a triangular tiling.  Phyllotactic state transitions, often accompanied by lattice defects, are observed when an incompatibility arises between a state and the confining geometry. Circularly symmetric phyllotaxis in constant curvature spaces leads to transitions characterised by circular grain boundaries~\cite{Sadoc2012, Sadoc2013}, as observed in sunflower heads~\cite{Jean1994}. Moreover, deviations from regular cylindrical confinement, by either varying the radius of the particles being packed~\cite{Zagorska2008} or the geometry of the system~\cite{Pennybacker2015, Atela2011,Beller2016, Mughal2017}, allow transitions between phyllotactic states. Understanding these transitions, as well as the resulting defect structures at the transitions, could give important insight into phyllotactic transitions in plants, such as are observed in some cacti~\cite{Shipman2005} and Agave Parryi~\cite{Rivier2016}, which is important for understanding plant growth. 

In this study, we numerically investigate the ground states of repulsive point-like particles, modelled as vortices in a type II superconductor, confined to the surface of a conical frustum. Recent work by two of the authors~\cite{Tomlinson2020} investigated the ground state structures of this system under cylindrical confinement. This study showed that the ground state triangular lattice structures can be classified as phyllotactic states. Which state is the ground state depends on both the density of particles \(\rho\) and the cylinder circumference \(c\) through the value of a dimensionless parameter \(\alpha(c,\rho)\). The phyllotactic ground state changes when the value of \(\alpha\) crosses a transition value. Away from these transition points, the value of \(\alpha\) uniquely determines the ground state. In the current investigation, we first determine the ground state structures on the conical frustum. For the ground state \(\rho\) is fixed but the value of \(c\), and therefore \(\alpha\), varies with position. We show that the local \(\alpha\) value within the system determines the locally preferred state and we observe transitions between phyllotactic states when the value of \(\alpha\) crosses a transition point. These transitions are mediated by topological defects in the lattice structure, the nature of which we classify according to a set of phenomenological rules. Having established that the value of \(\alpha\) controls the local behavior of the system, we use this to our advantage to create a state free from defects by introducing a density gradient to reduce variation in \(\alpha\) within the system. We find that on the conical surface there is still variation in the value of \(\alpha\), limiting the range of parameters for which this approach is possible. We therefore seek out a surface, the horn, for which \(\alpha\) can be kept constant and  the single state can be sustained for a wider parameter range. Finally, we discuss how the state of constant \(\alpha\) provides an example of a conformal crystal~\cite{Rothen1993}.

\section{\label{sec:description}System Model}
Vortices in type-II superconductors provide an excellent platform for studying soft lattice systems due to the large range of lattice parameters for which the lattice is stable, as well as the possibility of controlling the density with an external magnetic field. Here we numerically model a system of vortices which interact along the surface via the pairwise interaction potential \(V(r) = V_0 K_0(r/\lambda)\), where \(K_0\) is a modified Bessel function of the second kind, \(r\) is the separation between the vortex pair, measured along the surface, and \(\lambda\) is the London penetration depth. We set \(\lambda=1\) and the constant \(V_0=1\) throughout. The interaction potential is repulsive, with logarithmic divergence at small vortex separations and exponential decay in the large separation limit. We impose a cut off radius \(r_c = 5a_0\) to all simulations, where \(a_0\) is the average lattice parameter in the system.  

We create our conical frustum with its axis of revolution along the \(\vu{z}\) direction and give all system dimensions in terms of the scaled system length, which spans the range \(0<z<1\). The spatially varying circumference is \(c(z)=c_0 + \Delta c\,z\), where \(c_0\) is the circumference at \(z=0\) and \(\Delta c =c_1-c_0>0\) determines the slope of the surface. Positions on the surface are described by their coordinates \((z,\theta)\), where \(0\leq\theta<2\pi\) is an angular coordinate. We minimise the effect of hard boundaries at the ends of the system by attaching vortex reservoirs of fixed density to these regions and only analysing states at least a distance \(r_c\) from the hard boundaries. We utilise the zero Gaussian curvature of the conical frustum and consider the `unwrapped' surface in the plane, with boundary conditions identifying the two cut edges as equivalent, as shown in the schematic in Figure~\ref{fig:statictransition}(a).

Experimentally vortices can be confined to channels in thin film superconductors, the exact geometry of which controlled using nanofabrication techniques~\cite{Pruymboom1988, Theunissen1996, Papari2016}. A variant of our flattened system could therefore be fabricated and an appropriate choice of pinning array could be used to observe lattice states equivalent to those in our system. We note that the numerically observed ground states for cylindrical confinement are the same for the Bessel function, Yukawa and radially repulsive Gaussian potentials up to an overall energy scaling~\cite{Tomlinson2020}. We therefore expect that the results presented in this work are generic for point-like particles confined to the surfaces considered here for the case of repulsive pairwise interactions along the surface.

States on the surface are identified according to their phyllotactic notation, which gives a way of classifying the arrangement of the lattice structure on the surface. To determine the phyllotactic notation of a state, we assign each lattice site to a set of rows~\cite{Mughal2014}. A row is the set of vortices visited by hopping along the direction of one of the lattice vectors from a chosen starting point. Each vortex belongs to three rows, which correspond to the directions of the three lattice vectors of the triangular lattice. The phyllotactic notation for a state is written as three integers \((P,Q,R)\), which correspond to the number of rows of each orientation required to form the state. The three integers are written in descending order and satisfy \(P=Q+R\). The right side of Fig.~\ref{fig:statictransition}(c) shows the \((4,2,2)\) state with the lattice vectors \((\vb{a},\vb{b},\vb{c})\) labelled. The state consists of four rows of lattice parallel to the \(\vb{c}\) direction, and two rows parallel to each of the \(\vb{a}\) and \(\vb{b}\) directions. Sections of the two rows of lattice parallel to the \(\vb{b}\) direction are highlighted in light and dark blue and trace out spirals on the surface. All lattice rows forming a state span the length of the system except those parallel to the periodicity direction \(\vu{e}_{\theta}\), which form closed loops around the circumference at particular \(z\) values. These states have \(R=0\), such as the \((3,3,0)\) state on the left of Fig.~\ref{fig:statictransition}(c). All rows which are not parallel or perpendicular to \(\vu{e}_{\theta}\) trace out spirals on the surface.

\section{\label{sec:groundstates}Ground state structures on a conical frustum}

\begin{figure*}
		\vspace{-3em}
		\centering
		\includegraphics[width=0.9\textwidth]{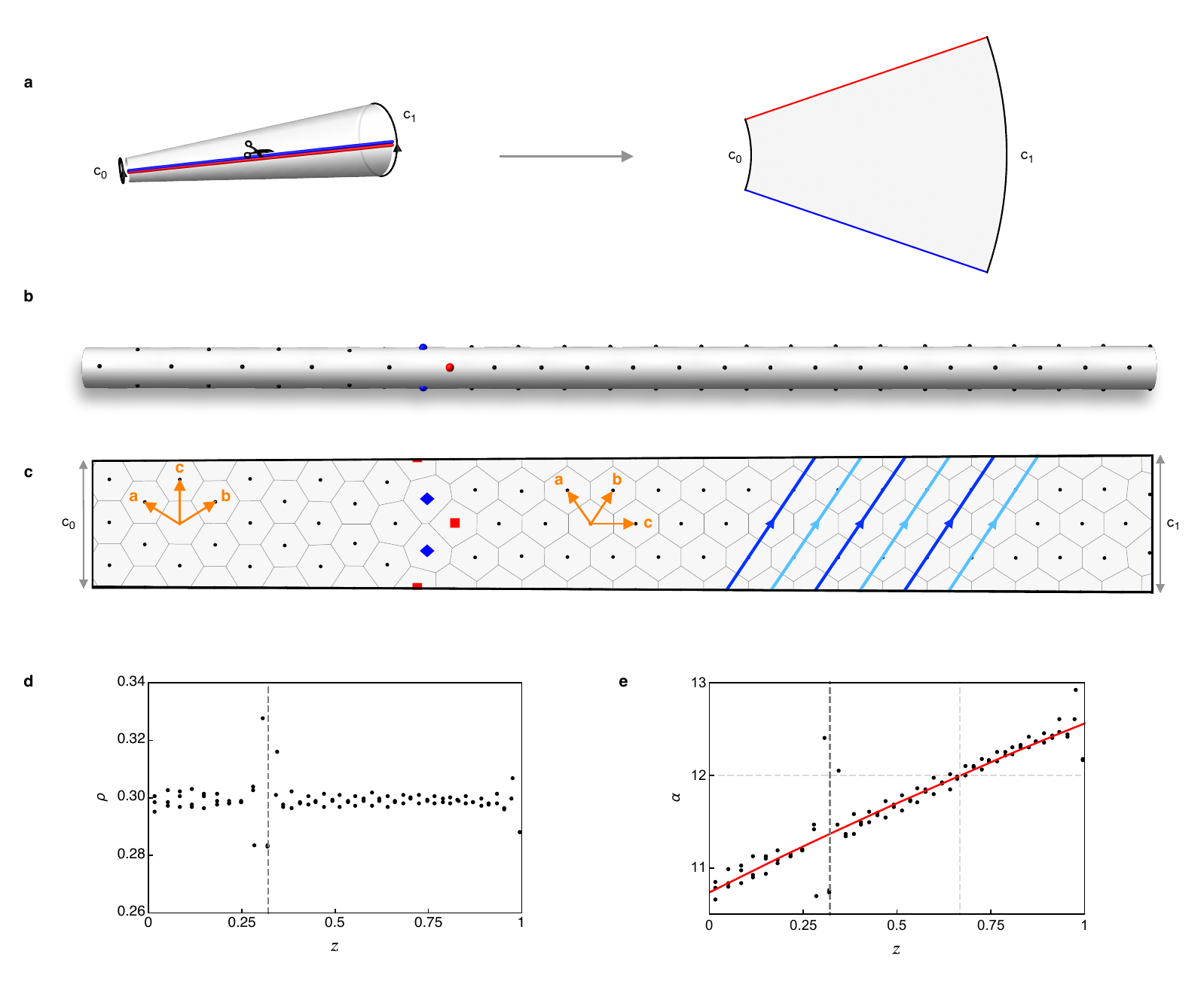}
	\caption{Ground state structures on a conical frustum show transitions between phyllotactic states. (a) Schematic of the conical system and cutting procedure to form the unwrapped shape. The red and blue edges are identified as equivalent. (b) System snapshot showing the transition between the (3,3,0) (left) and (4,2,2) (right) states on the conical surface. (c) Unwrapped version of the snapshot in (b) with Voronoi cells shown for all vortices. Dislocations, which mark the transition location, are pairs of vortices in which one has five (red square) and the other has seven (blue diamond) nearest neighbours. The lattice vectors for each state are labelled in orange. The alternating light and dark blue arrows on the right highlight the two rows of lattice parallel to the \(\vb{b}\) direction of the \((4,2,2)\) state. The number of rows of lattice along each of the three lattice directions determines the phyllotactic state. (d) Local density \(\rho\) for this transition plotted as a function of position \(z\). The vertical dashed line at \(z=0.32\) indicates the location of the transition, which also corresponds to a region with some local variation in the density. (e) Positional variation of \(\alpha=c^2\rho\), which dictates the preferred state. The dark dashed line at \(z=0.32\) marks the observed location of the transition while the light dashed lines mark the expected \(\alpha\) value at the transition (\(\alpha=12\)) and its corresponding position within the system (\(z=0.67\)). The measured \(\alpha\) value at the transition is 11.4.}
\label{fig:statictransition}
\end{figure*}

We first examine the ground states of the system by maintaining a fixed vortex density \(\rho\). The parameter \(\alpha\) is defined by \(\alpha=c^2\rho\), which is equivalent to the definition used in Ref.~\citenum{Tomlinson2020} for cylindrical confinement but which highlights the two dimensional nature of our system. We expect that the ground state structures of our system will depend on the local value of \(\alpha(z)=c^2(z)\rho\) and will be consistent with the ground state structures of the system under cylindrical confinement~\cite{Tomlinson2020} (Supplementary Tab. S1 online lists each of the states and their corresponding \(\alpha\) values). When the value of \(\alpha\) crosses a transition point, we expect to observe a state transition on the conical surface. 

We numerically search for the zero-temperature ground states using both molecular dynamics (MD) and Monte Carlo (MC) algorithms (see Methods). In the cylindrical system, the energy difference between some ground states and other higher-energy states is very small. We therefore use two methods for the conical system to ensure we have indeed found the ground states. We find that our results are consistent for the two methods, and provide an example comparing the two in Figs. S1 and S2 of the Supplementary Information online. We focus our investigation on the transition points between states and choose values of \(c_0\), \(\Delta c\) and \(\rho\) so that the resulting variation in \(\alpha\) leads to the value of \(\alpha\) crossing a single transition point at a position around \(z=1/2\). We note that under cylindrical confinement, some transitions involve a change of lattice vectors without the three-index phyllotactic notation changing. These transitions are not included in this investigation due to distortions of the states on the conical surface making the states too difficult to distinguish.

Using these techniques, we have observed all expected transitions between phyllotactic states which appear sequentially as \(\alpha\) is increased over the range \(4\leq\alpha\leq95.3\). As an example, the transition between the (3,3,0) and (4,2,2) states, which is predicted to occur at \(\alpha=12\) is shown in Fig.~\ref{fig:statictransition}(b)-(c), where Fig.~\ref{fig:statictransition}(b) shows the state on the conical surface while Fig.~\ref{fig:statictransition}(c) shows the flattened state. The system parameters are \(c_0=0.12\) and \(\Delta c = 0.01\). Transitions are mediated by lattice defects in the form of dislocations. These are vortex pairs in which one has five (red square) and the other has seven (blue diamond) nearest neighbours instead of the usual six neighbours (black circles). The Voronoi cells, which indicate the region closer to a given point than any other point, are shown for all vortices in Fig.~\ref{fig:statictransition}(c). We calculate the local density of each vortex as the inverse of its Voronoi cell area and use this to calculate its corresponding \(\alpha\) value. Plots of the local \(\rho\) and \(\alpha\) values for our example transition are shown in Fig.~\ref{fig:statictransition}(d) and (e) respectively. As expected, \(\rho\) is constant while \(\alpha\) increases with \(z\). Both plots show spreading in the data around the transition point. 

We find state transitions occur at \(\alpha\) values consistent with transition points for cylindrical confinement, although small differences between the expected and measured transition points are observed. In the example in Fig.~\ref{fig:statictransition}(d), the expected transition is at \(\alpha=12\) (horizontal dashed line) while the actual transition occurs at \(\alpha=11.4\). These deviations are likely due to the competing energetics of lattice strain and defect formation as well as the way in which the states form during the annealing process and finite-size effects in the \(z\) direction.

\begin{figure}
        \centering
		\includegraphics[width = 0.45\textwidth]{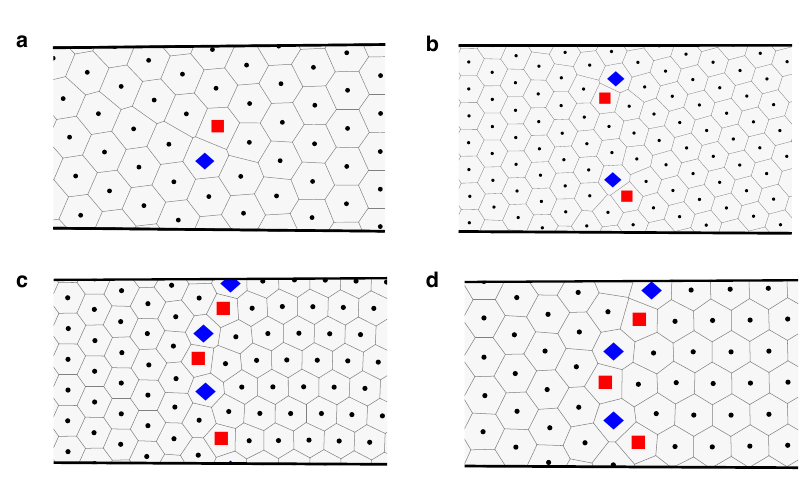}
		\caption{Examples of the different classes of defect patterns observed at transitions between states. (a) A single dislocation is present in the (5,4,1) to (5,5,0) transition. (b) Two separated dislocations mediate the (7,7,0) to (8,6,2) transition. (c) A grain boundary scar consisting of three dislocations is observed for the (6,6,0) to (7,4,3) transition. (d) A domain wall consisting of three dislocations forms for the (5,5,0) to (6,3,3) transition.}
	\label{fig:defects}
\end{figure}

At the transitions, we find that the number and arrangement of lattice defects depends on the relative orientation of the two states at either side of the transition point. We have observed defect patterns which can be classified into distinct classes: single dislocations, multiple isolated dislocations, grain boundary scars, and domain walls. In the case of one or more isolated dislocations, as shown in Fig.~\ref{fig:defects}(a)-(b) for the (5,4,1) to (5,5,0) and (7,7,0) to (8,6,2) transitions respectively, the defects are completely surrounded by vortices with six nearest neighbours. Grain boundary scars are lines of two or more adjacent dislocations at the transition, an example of which is shown in Fig.~\ref{fig:defects}(c) for the (6,6,0) to (7,4,3) transition. Finally, a domain wall, in which a line of dislocations spans the entire circumference, is shown in Fig.~\ref{fig:defects}(d) for the (5,5,0) to (6,3,3) transition.

We observe that domain walls form when one state at the transition has one of its lattice vector parallel \(\vu{e}_{\theta}\) while the other state has a lattice perpendicular to this. States with these properties have the general notation \((Q,Q,0)\) and \((2R,R,R)\) respectively. Grain boundary scars are observed when the transition is between states with general notation \((Q,Q,0)\) and \((2R+1,R+1,R)\). The \((2R+1, R+1, R)\) states have a lattice vector with a large component perpendicular to \(\vu{e}_{\theta}\). In both of these cases, the competing structures are incompatible so the transition requires defects to span much or all of the circumference.

Furthermore, can predict the number of dislocations \(n_d\) at a transition. Dislocations in a lattice structure are signatures of a change in the number of rows of lattice in different regions. We observe \(n_d = |\Delta R|\), where \(\Delta R\) is the difference between the smallest indices of the two phyllotactic states either side of the transition. The \(R\) index of the phyllotactic state corresponds to the number of rows along the direction of the lattice vector with the largest \(\vu{e}_{\theta}\) component. Due to the commensurability constraint between the periodicity of the lattice structure and circumference of the surface, it is easier for the lattice to distort 
perpendicular to \(\vu{e}_{\theta}\) than along it. The \(R\) index therefore corresponds to the most confined set of rows forming the state and it is the change in this number of rows between the competing states which determines the value of \(n_d\).

\section{\label{sec:coneflow}Maintaining a single lattice structure with a density-driven flow}
We now turn our attention to designing a vortex configuration on the conical surface which is free from topological defects. We have established that the ground state behaviour of our system is dependent on the local value of the parameter \(\alpha\), which varies due to the changing system circumference. In a type-II superconductor, the vortex density is directly proportional to the magnitude of the applied external magnetic field. We therefore introduce a density gradient, which is equivalent to a magnetic field gradient, to counteract the increasing circumference and reduce variation in \(\alpha\). In this section, we demonstrate that it is possible to create a defect-free state using this approach and discuss the limitations.

\begin{figure*}
\centering
\includegraphics[width=\textwidth]{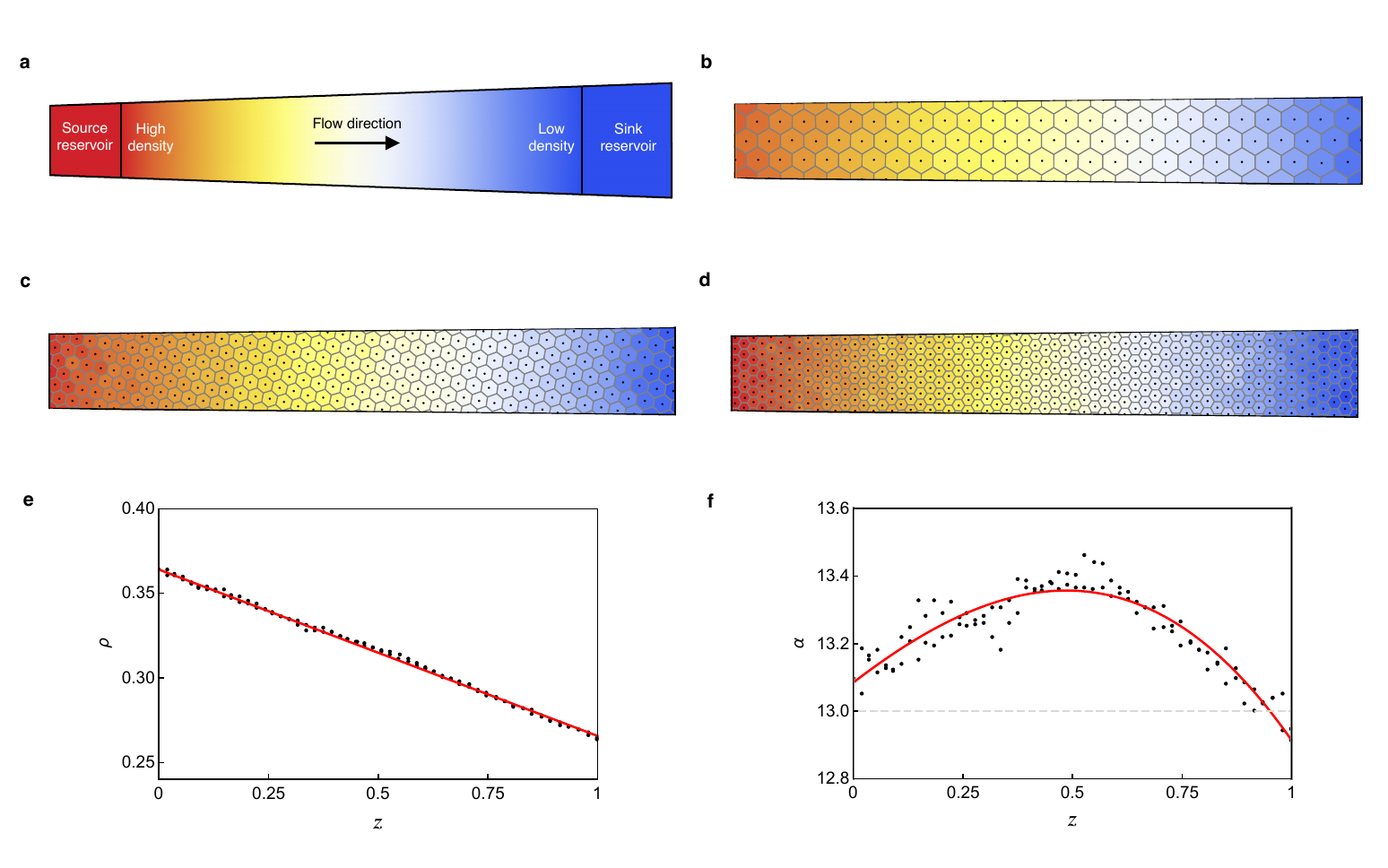}
\caption{Maintaining a single state by inducing a density-driven flow. (a) Schematic of the source-sink density-driven flow scheme. (b) (4,2,2) state with target value \(\alpha_0 = 13\), (c) (6,4,2) state with \(\alpha_0 = 32\), and (d) (7,7,0) state with \(\alpha_0 = 57\). In (b)-(d) the colour gradient indicates the density from highest (red) to lowest (blue). The flow is towards the region of lowest density. (e) Local density variation within the system as a function of distance \(z\). This corresponds to the snapshot in (b) with a linear fit line (red) plotted. (f) Local variation in \(\alpha\) along the system, corresponding to the snapshot in (b). The red line gives a cubic fit for the local \(\alpha\) value.}
\label{fig:densitydriven}
\end{figure*}

Using MD, we create a density gradient \(\rho(z)\) by implementing a source-sink density-driven flow scheme. This is achieved by attaching reservoirs maintained at fixed vortex number to the ends of system, following the methods detailed in Ref.~\citenum{Watkins2019} (also see Methods). A vortex flow is induced along the \(z\) direction towards the reservoir maintained at the lower density (see schematic, Fig.~\ref{fig:densitydriven}(a)). A target value \(\alpha_0\) is chosen for vortices entering and exiting the system. This value is chosen to be away from transition values of \(\alpha\) to allow the reservoirs to maintain the desired state. The required vortex densities \(\rho_0\) at \(z=0\) and \(\rho_1\) at \(z=1\) are calculated from the value of \(\alpha_0\) and used to determine the number of vortices in each reservoir. 

Using the parameters \(c_0=0.12\) and \(\Delta c = 0.02\) and sampling a range of \(\alpha_0\) values, we have observed all of the expected configurations which are successive ground states on the cylinder from (2,2,1) through to (7,7,0), as listed in Supplementary Tab. S1 online. After steady state is reached, which typically takes around \(10^5\) time steps, a single state is maintained throughout the system provided that the variation in \(\alpha\) is small enough for the state to be maintained. For higher \(\alpha\) values, the states become more unstable to fluctuations in the system and are much more difficult to maintain. Figure~\ref{fig:densitydriven}(b)-(d) show snapshots of the density-driven flow scheme for three examples: (b) the (4,2,2) state in a system with \(\alpha_0 = 13\), (c) the (6,4,2) state with \(\alpha_0 = 32\), and (d) the (7,7,0) state with \(\alpha_0 = 57\). In each case, the colour gradient indicates the density, from highest (red) to lowest (blue). The direction of flow is from left to right. We observe that the relative orientation between the lattice vectors and the flow direction does not appear to affect the behaviour. 

While the steady state flow shows no defects, there is still underlying variation in \(\alpha\) within the system which restricts the parameter range for which a single state can be maintained. The implementation of the density gradient leads to a linear density profile of the form \(\rho(z) = \rho_0+\Delta \rho z\), where \(\Delta\rho = \rho_1-\rho_0 <0\), as shown in the graph in Fig.~\ref{fig:densitydriven}(e). Figure~\ref{fig:densitydriven}(f) shows the resulting variation in \(\alpha(z)=c^2(z)\rho(z)\), which is therefore cubic in \(z\). For small enough \(\Delta c\) and \(\Delta \rho\) values it is possible to maintain the system within a single state.  However, even when the single state is maintained, the structure is frustrated and the vortices are not arranged as would be expected if \(\alpha\) were constant. Figure~\ref{fig:horn}(a) shows 
the deviation between expected (orange) and actual (black) vortex positions in a system snapshot of the (4,2,2) state with \(c_0=0.126\) and \(\Delta c=0.064\). We find this state is able to accommodate larger \(\alpha\) variation, and therefore larger magnitude \(\Delta c\) and \(\Delta \rho\) values, than many of the other states and can therefore be used to highlight deviations from a constant \(\alpha\) state. In the next section we refine our system to create a setup for which the value of \(\alpha\) can be held constant.

\section{Surface with constant \(\alpha\)}
In order to design a surface with constant \(\alpha\), we must create a system for which we can balance the changing circumference of the surface and the changing density profile. The implementation of the density-driven flow scheme gives us a linear density profile and we therefore solve for the profile \(c(z)\) which gives a constant value of \(\alpha = \alpha_0\). This leads to a horn-shaped surface with 
	\begin{align}
		c(z) = \sqrt{\frac{\alpha_0}{\rho_0+\Delta\rho z}} \equiv \frac{c_0}{\sqrt{1-\eta z}},
	\end{align}	
where \(c_0\) is the circumference at \(z=0\), \(\eta = -\Delta\rho/\rho_0 >0\) and the length of the system is scaled such that \(0<z<1\).

\begin{figure*}
\centering
\includegraphics[width=0.9\textwidth]{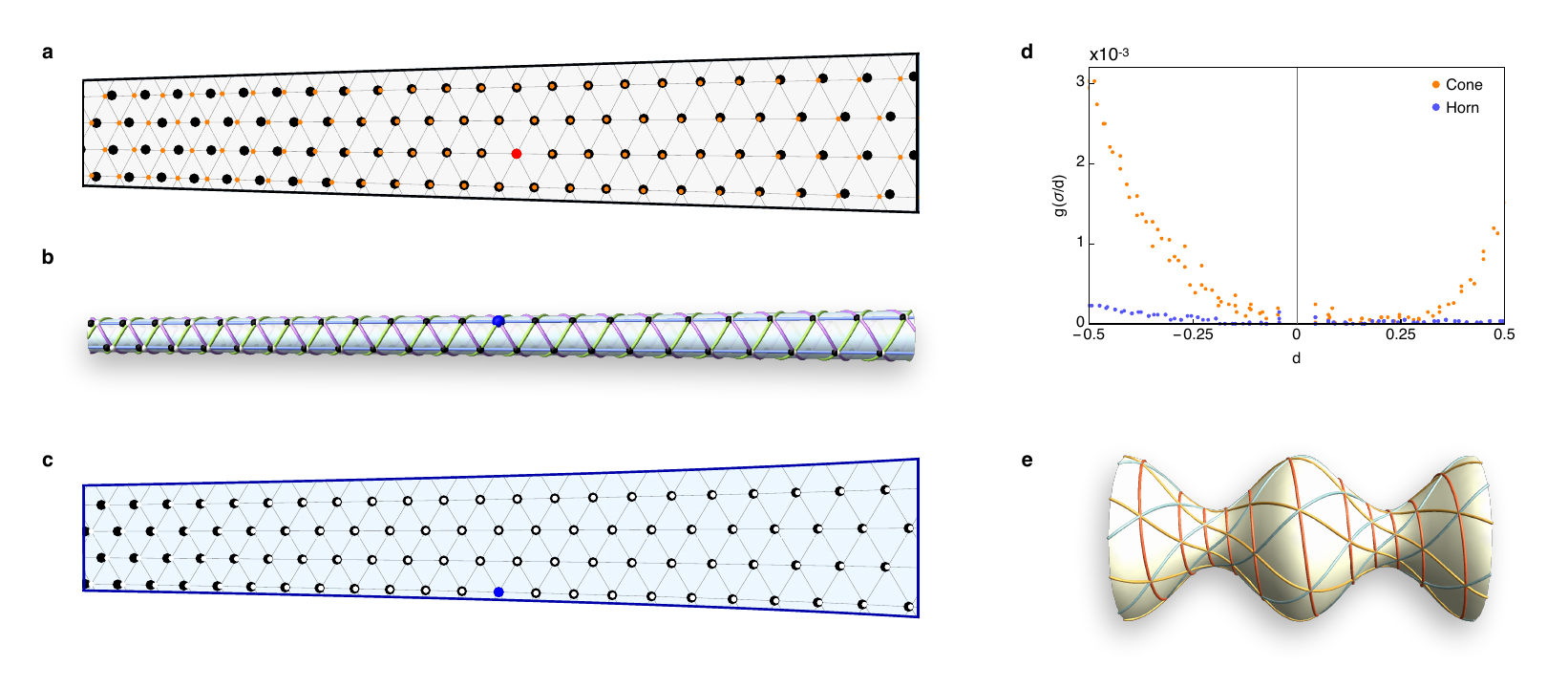}
\caption{Comparison between cone and horn systems. (a) Snapshot of the flattened cone in the (4,2,2) state showing the actual vortex positions (black) and expected vortex locations (orange) if \(\alpha\) were constant. The expected positions are calculated relative to the red reference point. (b)-(c) Snapshot of (b) the horn surface and (c) unwrapped surface showing the (4,2,2) state with comparable system parameters to the cone example in (a). Actual vortex locations are shown in black. The expected vortex locations (white) are calculated relative to the blue reference point. These sit at the intersections of the plotted loxodrome curves. (d) Plot showing the deviation between the expected and actual vortex positions \(g(\sigma/d)\) (defined in the main text) as a function of distance from the reference point \(d\). The plot corresponds to the images in (a)-(c) for the cone (orange) and horn (blue) respectively. (e) (7,6,1) state of constant \(\alpha\) constructed on the surface formed by revolving the curve \(y(x) = \cos{x}+2\) about the \(x\) axis.}
\label{fig:horn}
\end{figure*}

We perform MD on the surface by numerically solving for the geodesics connecting pairs of vortices. The distance between a pair is the length of the geodesic and the forces act along the tangent vectors to the geodesics. As per the conical system, we attach source and sink reservoirs to maintain the required density gradient. Details of the equations and techniques used for MD in this case are given in the Supplementary Information online. We choose the parameters \(c_0 = 0.126\) and \(\eta=0.56\) so that the circumferences at \(z=0\) and \(z=1\) are equivalent to those for the cone example in Fig.~\ref{fig:horn}(a). 

To verify our results, we need to calculate the expected vortex locations for a state of constant \(\alpha\) on the surface. The value of \(\alpha\) uniquely determines the direction of the lattice vectors used to describe a state, measured relative to the periodicity direction \(\vu{e}_{\theta}\). The magnitude of the lattice vectors depends on the local value of the circumference. Since the value of \(\alpha\) is constant on the surface, the lattice vectors have the same orientation relative to \(\vu{e}_{\theta}\) everywhere on the surface. From a given starting point, tracing out a curve by following the direction of a single lattice vector therefore creates a curve whose tangent vector is at a constant angle relative to \(\vu{e}_{\theta}\). This type of curve is known as a loxodrome and has applications in navigation as well as other physical systems~\cite{Alexander2004, Ansell2019} (see Supplementary Information online for derivation). By tracing out sets of these curves along the three lattice directions, starting from an initial reference point, we can create a series of curves which are equivalent to the rows forming the lattice state. The intersection points of the sets of curves give the expected vortex locations.

Fig.~\ref{fig:horn}(b)-(c) respectively show snapshots of the horn surface and the same snapshot flattened onto the plane. The vortices (black circles) are expected to sit at the intersection points of the plotted curves (white circles). Comparing results for the cone (Fig.~\ref{fig:horn}(a)) and horn (Fig.~\ref{fig:horn}(c)), we observe that the locations of the vortices for the horn system are significantly closer to their expected locations than those on the cone. We quantify the deviation between the expected and actual vortex locations for each system in Fig.~\ref{fig:horn}(d) by plotting \(g(\sigma/d) = \sigma^2/d^2\), where \(\sigma\) is the distance between a vortex and its expected location and \(d\) is the distance between the vortex and the reference point. For the horn (blue points) we observe that \(g(\sigma/d)\) is very close to zero, as expected for a constant \(\alpha\) system, while for the cone (orange points) there is a clear upward trend in \(g(\sigma/d)\) as \(d\) increases. We therefore conclude that the horn system is displaying a state of constant \(\alpha\).

Using the horn system for simulation is limited by the density variation within the system, and therefore the variation in the horn radius, due to the vortex interaction strength. If the variation is too large, vortices in the dense region will interact very strongly while those in the least dense regions may be far enough apart that the interactions are almost negligible.  There are also large computational costs in simulations due to the time required to calculate distances along the curved surface. 

\section{Conclusions}
We have demonstrated the adaptability of the vortex lattice in both static and dynamic regimes when confined to a geometry that is not commensurate with the local lattice ground state, as well as the importance of the dimensionless quantity \(\alpha\) in controlling this behaviour. By fixing the density, the ground state displays transitions between phyllotactic states as \(\alpha\) crosses a transition value. The number of defects at the transition can be described by a phenomenological rule. By inducing a density gradient in opposition to the changing circumference, we have created defect-free structures in which the entire system is in a single state. However, on the conical surface \(\alpha\) still varies so this approach is limited to small \(\Delta c\) and \(\Delta\rho\). We therefore sought a surface, the horn, for which the value of \(\alpha\) could be kept constant with our chosen density profile and demonstrated that in this case the vortices flow in a state of constant \(\alpha\). 

More generally, we note that the method presented for using loxodromes to construct a constant \(\alpha\) configuration can be applied to other surfaces of revolution, such as in the example in Fig.~\ref{fig:horn}(e) showing the (7,6,1) state on the surface of revolution created by revolving the curve \(y(x) = \cos{x}+2\) about the \(x\) axis. For surfaces for which the shape profile \(y(x)>0\) for the range of \(x\) values, all of the intersection points of curves forming states constructed using this method are six-fold coordinated. These states can all be conformally mapped from the surface to a perfect crystal structure in a strip of the plane~\cite{Alexander2004, Feeman2001}. They are therefore a different set of examples of conformal crystals, the classic examples of which describe the structural arrangement of repulsive particles under the influence of gravity~\cite{Rothen1993, Ray2013}.

\section*{Methods}
\subsection*{Ground state structures on a conical frustum}
We use Molecular dynamics (MD) methods throughout this investigation and additionally use Monte Carlo (MC) techniques to verify that we have indeed found the ground states. MD is performed by numerically solving the over-damped Langevin equation for the motion of each vortex~\cite{Jensen2003, Reichhardt2001} up to the imposed cut-off radius \(r_c=5a_0\), where \(a_0\) is the lattice parameter. We control temperature through implementation of a thermostat \cite{Jensen2003, Dong1993} and slowly anneal a pseudo-randomly generated initial configuration from a high temperature into the zero-temperature ground state. We simulate the flattened system with length \(L=50\), \(c_0 = 6.0\) and \(\Delta c = 0.5\) as the typical values. These parameters allow for enough variation in \(\alpha\) that there is a single transition point within the system. For higher \(\alpha\) transitions, typically \(\alpha>50\), we instead use \(\Delta c = 0.1\) and \(L = 30\) due to the higher density of the lattices involved in the transitions. Other choices of \(c_0\) and \(\Delta c\) have been investigated to verify they are consistent with the results presented here. Using this method we have investigated a range of densities large enough to observe all transitions in the range \(4\leq\alpha\leq95.3\).

We follow the Metropolis prescription for MC integration~\cite{Metropolis1953} by annealing a pseudo-randomly generated starting vortex configuration from an initially high temperature down to zero temperature. In each step, we propose a position update, selected from a normal distribution, that displaces a pseudo-randomly selected vortex by no more than \(\sqrt{2}L/500\). Temperature \(T\) is reduced linearly from \(T=0.001\) to zero over the course of the simulation, which ends after \(10^7\) proposed updates. The initial temperature is sufficiently high to allow vortices to effectively explore the energetic landscape to enable cooling towards the structural ground state. Snapshots of the systems are compared and the state with the lowest internal energy is selected as the numerically solved ground state. We perform MC minimisation for system sizes in the range \(4.0\leq c_0\leq 10.5\) with  \(\Delta c=0.5\) and \(L=50\), where we increment the value of \(c_0\) by 0.5 over the given range. For each system size, we choose the number of vortices \(N_v\) and find the ground state for each integer value in the range \(100\leq N_v \leq 256\). Using this range of parameters allows us to extensively explore the phase space for states expected in the range \(4\leq\alpha\leq42.1\). The results are consistent with those for MD and show the expected transitions when \(\alpha\) crosses a transition value within the system. The MC sampling techniques do not choose parameters such that a transition is expected to occur at the centre of the system. Rather, as \(N_v\) is increased for fixed system size, the observed position of a transition moves through the system. We use these results to verify our MD results, but focus our analysis on the parameter choices for which transitions occur at the centre of the channel and are therefore as far away from the boundaries as possible. \\

\subsection*{Density-driven flow on a conical frustum}
The vortex reservoirs are created to be larger than \(r_c\) and all vortices are added and removed from the source and sink respectively near to the ends of the reservoirs in order to ensure there are minimal effects on the dynamics within the system. Vortices are added into the source reservoir at pseudo-random locations in a region close to the outside edge of the system.  The vortices within this reservoir form into a particular state which then flows into the system. As the state is formed within the source, there are occasions where defects may form in the structure and escape into the system. This noise flows through the system and the expected state is the persistent structure. 

If a clean state entirely free from the possibility of defect structures entering from the source is desired for a different application, steps can be taken to produce a more ordered source reservoir. Positioning the vortices at their expected positions for the state of interest within the source helps to eliminate the chance of defects persisting and escaping the source.

\subsection*{Density-driven flow on a horn-shaped surface}
We follow the methods used on the conical system as closely as possible, making adaptations to account for the curved surface where necessary. As for the conical system, we impose a cut-off radius of \(5a_0\) for interactions. The density gradient means that the separation between vortices varies with position. We therefore choose the value of \(a_0\) to be the expected lattice parameter at \(z=0.5\). Details of the methodology used for molecular dynamics on the horn surface are given in the Supplementary Information online.


\section*{Acknowledgements}
The authors would like to thank R. D. Kamien, M. Dennis, J. Watkins, R. Mason, J. M. F. Gunn and C. Wilkin for their valuable insight and discussions. H. S. A. was supported by NSF MRSEC Grant No. DMR-1720530 and a Simons Investigator Grant from the Simons Foundation to Randall D. Kamien. A.A.T. received financial support from the EPSRC for student funding.

\section*{Author contributions}
H.S.A., A.A.T. and N.K.W. designed the investigation. H.S.A. and A.A.T. performed simulations. H.S.A., A.A.T. and N.K.W. analysed the results and wrote the manuscript.

\section*{Competing interests}
The authors declare no competing interests.


\begin{thebibliography}{10}

\bibitem{Giomi2008a}
Giomi, L. \& Bowick, M. J. 
Defective ground states of toroidal crystals.
{\em Phys. Rev. E} {\bf 78,} 010601(R) (2008).

\bibitem{Giomi2008b}
Giomi, L. \& Bowick, M. J. 
Elastic theory of defects in toroidal crystals.
{\em Eur. Phys. J. E} {\bf 27,} 275--296 (2008).

\bibitem{Bausch2003}
Bausch, A. R. {\em et. al.}
Grain boundary scars and spherical crystallography.
{\em Science} {\bf 299,} 1716--1718 (2003).

\bibitem{Lipowsky2005}
Lipowsky, P., Bowick, M. J., Meinke, J. H., Nelson, D. R. \& Bausch, A. R.
Direct visualization of dislocation dynamics in grain-boundary scars.
{\em Nat. Mater.} {\bf 4} 407--411 (2005).

\bibitem{Sausset2010}
Sausset, F., Tarjus, G. \& Nelson, D. R.
Structure and dynamics of topological defects in a glassy liquid on a negatively curved manifold.
{\em Phys. Rev. E} {\bf 81,} 031504 (2010).

\bibitem{Einert2005}
Einert, T., Lipowsky, P., Schilling, J., Bowick, M. J. \& Bausch. A. R.
Grain boundary scars on spherical crystals.
{\em Langmuir} {\bf 21,} 12076--12079 (2005).

\bibitem{Burke2015}
Burke, C. J., Mbanga, B. L., Wei, Z., Spicer, P. T. \& Atherton, T. J.
The role of curvature anisotropy in the ordering of spheres on an ellipsoid.
{\em Soft Matter} {\bf 11,} 5872--5882 (2015).

\bibitem{Soni2018}
Soni, V., G{\'{o}}mez, L. R. \& Irvine, W. T. M.
Emergent geometry of inhomogeneous planar crystals.
{\em Phys. Rev. X} {\bf 8,} 011039 (2018).

\bibitem{Irvine2010}
Irvine, W. T. M., Vitelli, V. \& Chaikin, P. M.
Pleats in crystals on curved surfaces.
{\em Nature} {\bf 468,} 947--951 (2010).

\bibitem{Mughal2014}
Mughal, A. \& Weaire, D.
Theory of cylindrical dense packings of disks.
{\em Phys. Rev. E} {\bf 89,} 042307 (2014).

\bibitem{Erickson1973}
Erickson, R. O.
Tubular packing of spheres in biological fine structure.
{\em Science} {\bf 181,} 705--716 (1973).

\bibitem{Pickett2000}
Pickett, G. T., Gross, M. \& Okuyama, H.
Spontaneous chirality in simple systems.
{\em Phys. Rev. Lett.} {\bf 85,} 3652--3655 (2000).

\bibitem{Lohr2010}
Lohr, M. A. {\em et. al.}
Helical packings and phase transformations of soft spheres in cylinders.
{\em Phys. Rev. E} {\bf 81,} 040401(R) (2010).

\bibitem{Mughal2012}
Mughal, A., Chan, H. K., Weaire, D. \& Hutzler, S.
Dense packings of spheres in cylinders: simulations.
{\em Phys. Rev. E} {\bf 85,} 051305 (2012).

\bibitem{Fu2016}
Fu, L., Steinhardt, W., Zhao, H., Socolar, J. E. S. \& Charbonneau, P.
Hard sphere packings within cylinders.
{\em Soft Matter} {\bf 12}, 2505--2514 (2016).

\bibitem{Kholbystov2004}
Khlobystov, A. N., Britz, D. A., Ardavan, A. \&. Briggs, G. A. D.
Observation of ordered phases of fullerenes in carbon nanotubes.
{\em Phys. Rev. Lett.} {\bf 92,} 245507 (2004).

\bibitem{Yamazaki2008}
Yamazaki, T. {\em et. al.}
Ordered fullerene nanocylinders in large-diameter carbon nanotubes.
{\em Nanotechnology} {\bf 19,} 045702 (2008).

\bibitem{Liang2014}
Liang, R. {\em et. al.}
Assembly of polymer-tethered gold nanoparticles under cylindrical confinement.
{\em ACS Macro Lett.} {\bf 3,} 486--490 (2014).

\bibitem{Oguz2011}
O\u{g}uz, E. C., Messina,  R. \& L\"{o}wen, H.
Helicity in cylindrically confined Yukawa systems.
{\em EPL} {\bf 94,} 28005 (2011).

\bibitem{Tomlinson2020}
Tomlinson, A.~A. \& Wilkin, N. K.
Controlled transitions between phyllotactic states of repulsive
  particles confined on the surface of a cylinder.
{arXiv: 2001.03948 [cond-mat.soft]} (2020).

\bibitem{Sadoc2012}
Sadoc, J.-F., Rivier, N. \& Charvolin, J.
Phyllotaxis: a non-conventional crystalline solution to packing efficiency in situations with radial symmetry.
{\em Acta Cryst.} {\bf 68} 470--483 (2012).

\bibitem{Sadoc2013}
Sadoc, J. F., Charvolin, J. \& Rivier, N.
Phyllotaxis on surfaces of constant Gaussian curvature.
{\em J. Phys. A Math. Theor.} {\bf 46,} 295202 (2013).

\bibitem{Jean1994}
Jean, R. V.
{\em Phyllotaxis: A Systematic Study In Plant Morphogenesis}.
(Cambridge University Press, 1994).

\bibitem{Zagorska2008}
Zag{\'{o}}rska-Marek, B \& Szpak, M.
Virtual phyllotaxis and real plant model cases.
{\em Funct. Plant Biol.} {\bf 35,} 1025--1033 (2008).

\bibitem{Pennybacker2015}
Pennybacker, M. F., Shipman, P. D. \& Newell, A. C.
Phyllotaxis: some progress, but a story far from over.
{\em Physica D} {\bf 306,} 48--81 (2015).

\bibitem{Atela2011}
Atela, P.
The geometric and dynamic essence of phyllotaxis.
{\em Math. Model Nat. Phenom.} {\bf 6,} 173--186 (2011).

\bibitem{Beller2016}
Beller, D. A. \& Nelson, D. R.
Plastic deformation of tubular crystals by dislocation glide.
{\em Phys. Rev. E} {\bf 94,} 033004 (2016).

\bibitem{Mughal2017}
Mughal, A. \& Weaire, D.
Phyllotaxis, disk packing, and Fibonacci numbers.
{\em Phys. Rev. E} {\bf 95,} 022401 (2017).

\bibitem{Shipman2005}
Shipman, P. D. \& Newell, A. C.
Polygonal planforms and phyllotaxis on plants.
{\em J. Theor. Biol.} {\bf 236,} 154--197 (2005).

\bibitem{Rivier2016}
Rivier, N.,  Sadoc, J.-F. \& Charvolin, J.
Phyllotaxis: a framework for foam topological evolution.
{\em Eur. Phys. J. E} {\bf 39}, 7 (2016).

\bibitem{Rothen1993}
Rothen, F., Pieranski, P., Rivier, N. \& Joyet, A. 
Conformal crystal.
{\em Eur. J. Phys} {\bf 14,} 227 (1993).

\bibitem{Pruymboom1988}
Pruymboom, A., Kes, P. H., van der Drift, E. \& Radelaar, S.
Flux-line shear through narrow constraints in superconducting films.
{\em Phys. Rev. Lett.} {\bf 60,} 1430 (1988).

\bibitem{Theunissen1996}
Theunissen, M. H., Van der Drift, E. \& Kes, P. H.
Size effects in flow of flux-line solids and liquids.
{\em Phys. Rev. Lett.} {\bf 77,} 159 (1996).

\bibitem{Papari2016}
Papari, G. P.  {\em et. al.}
Geometrical vortex lattice pinning and melting in YBaCuO submicron bridges.
{\em Sci. Rep.} {\bf 6,} 38677 (2016).

\bibitem{Watkins2019}
Watkins, J. S. \& Wilkin, N. K.
Extruding the vortex lattice: two reacting populations of dislocations.
{\em EPL} {\bf 126,} 16002 (2019).

\bibitem{Alexander2004}
Alexander, J.
Loxodromes: a rhumb way to go.
{\em Math. Mag} {\bf 77,} 349--356 (2004)

\bibitem{Ansell2019}
Ansell, H. S., Kim, D. S., Kamien, R. D., Katifori, E. \& Lopez-Leon, T.
Threading the spindle: a geometric study of chiral liquid crystal polymer microparticles.
{\em Phys. Rev. Lett.} {\bf 123,} 157801 (2019).

\bibitem{Feeman2001}
Feeman, T. G.
Conformality, the exponential function, and world map projections.
{\em Coll. Math. J.} {\bf 32,} 334-342 (2001).

\bibitem{Ray2013}
Ray, D., Olson Reichhardt, C. J., Jank\'{o}, B. \& Reichhardt, C.
Strongly enhanced pinning of magnetic vortices in type-II superconductors by conformal crystal arrays.
{\em Phys. Rev. Lett.} {\bf 110,} 267001 (2013).

\bibitem{Jensen2003}
Jensen, H. J.
Simulations of relaxation, pinning and melting in flux lattices.
In {\em Phase Transitions and Relaxation in Systems with Competing
  Energy Scales} (ed. Riste, T. \& Sherrington, D.), 129--185 (Kluwer Academic Publ., 1993).

\bibitem{Reichhardt2001}
Reichhardt, C., Olson, C. J., Scalettar, R. T. \& Zim{\'{a}}nyi, G. T.
Commensurate and incommensurate vortex lattice melting in periodic pinning arrays.
{\em Phys. Rev. B} {\bf 64,} 144509 (2001).

\bibitem{Dong1993}
Dong, J.
Simulation of the vortex motion in high-T$_c$ superconductors.
{\em J. Phys. Condens. Matter} {\bf 5,} 3359 (1993).

\bibitem{Metropolis1953}
Metropolis, N., Rosenbluth, A. W., Rosenbluth, M. N. \& Teller, A. H.
Equation of state calculations by fast computing machines.
{\em J. Chem. Phys.} {\bf 21,} 1087 (1953).


\end{thebibliography}
\end{document}


\flushbottom

\maketitle

\renewcommand{\theequation}{\thesection.\arabic{equation}}
\renewcommand{\thefigure}{S\arabic{figure}}
\renewcommand{\thetable}{S\arabic{table}}
\numberwithin{equation}{section}

\newpage
\appendix


\section{\(\alpha\) values at transitions}

Table \ref{tab:transitions} states the transition values \(\alpha_T\) at which the lattice state transitions into the given phyllotactic state as the value of \(\alpha\) is increased from \(\alpha=0\). The transition values are calculated for the case of cylindrical confinement. This table is a reproduction of results presented previously in Ref.~[20] of the main text, extended here to \(\alpha = 100.824\). 
\begin{center}
\begin{table}[h]
\centering
\begin{tabular}{|c|c|c|c|c|c|c|c|}
\cline{1-2}\cline{4-5}\cline{7-8}
\(\quad\quad\alpha_T\quad\quad\) &\(\quad\)State\(\quad\)  & & \(\quad\quad\alpha_T\quad\quad\) & \(\quad\)State\(\quad\)& & \(\quad\quad\alpha_T\quad\quad\) &\(\quad\)State\(\quad\) \\ \cline{1-2}\cline{4-5}\cline{7-8}
0 & (1, 1, 0) & & 30 & (6, 3, 3) & & 68.0683 & (9, 5, 4) \\ \cline{1-2}\cline{4-5}\cline{7-8}
2 & (2, 1, 1) & & 31.7613 & (6, 4, 2) & & 71.604 & (9, 6, 3)\\ \cline{1-2}\cline{4-5}\cline{7-8}
4 & (2, 2, 0)  & & 34.0042 & (6, 5, 1) & & 73.3 & (8, 8, 0)\\ \cline{1-2}\cline{4-5}\cline{7-8}
\(8\sqrt{3/5}\) &(3, 2, 1) & & 38.7034 & (6, 6, 0)& & 75.7325 & (9, 7, 2) \\ \cline{1-2}\cline{4-5}\cline{7-8}
9.1912 & (3, 3, 0)& & \(504/\sqrt{143}\) & (7, 4, 3)& & 80.7751 & (9, 8, 1) \\ \cline{1-2}\cline{4-5}\cline{7-8}
12 & (4, 2, 2)  & & 43.9001 & (7, 5, 2) & & 85.4282 & (10, 5, 5) \\ \cline{1-2}\cline{4-5}\cline{7-8}
14.4506 & (4, 3, 1)  & & 47.2565 & (7, 6, 1) & & 87.1823 & (10, 6, 4) \\ \cline{1-2}\cline{4-5}\cline{7-8}
16.7283 & (4, 4, 0)  & & \(416/3\sqrt{7}\) & (8, 4, 4) & & 89.5129 & (10, 7, 3) \\ \cline{1-2}\cline{4-5}\cline{7-8}
\(160/3\sqrt{7}\) & (5, 3, 2) & & 56 & (7, 7, 0)& & 92.3081 & (9, 9, 0) \\ \cline{1-2}\cline{4-5}\cline{7-8}
23.0956 & (5, 4, 1) & & 58.4359 & (8, 6, 2)& & 95.341 & (10, 8, 2) \\ \cline{1-2}\cline{4-5}\cline{7-8}
26.5641 & (5, 5, 0) & & 62.8623 & (8, 7, 1) & &100.824 & (11, 6, 5) \\ \cline{1-2}\cline{4-5}\cline{7-8}
\end{tabular}
\caption{Transition values \(\alpha_T\) for which the system adopts the given phyllotactic state as the value of \(\alpha\) is increased.}
\label{tab:transitions}
\end{table}
\end{center}

\section{Comparison between molecular dynamics and Monte Carlo results}
In this section we give a direct comparison between results obtained for ground states using molecular dynamics (MD) and Monte Carlo (MC) methods. We use as our example the transition between the \((4,3,1)\) and \((4,4,0)\) states, which is expected to occur at \(\alpha=16.73\). Figures S1(a) and S2(a) show snapshots of the ground states obtained using MD and MC respectively. In each snapshot there are 130 vortices and the system dimensions are \(L=50\), \(c_0=6.0\) and \(\Delta c=0.5\).

In Figs. S1(b) and S2(b) we plot the density variation within each system, while in Figs. S1(c) and S2(c) we plot the resulting variation in the parameter \(\alpha\). The dark grey dashed lines in each graph correspond to the position of the transition within the system, which is marked by the presence of a single dislocation in the system snapshots. The light grey lines on each plot represent the expected \(\alpha\) value of the transition and the corresponding position within the system at which the local value of \(\alpha\) takes on that value, which is calculated from the red best-fit curve. In both cases, the difference between the expected and actual positions of the transition within the system corresponds to less than 4\% of the system length. This difference is the length of the average lattice parameter in the snapshots.

	\begin{figure}[!htb]
		\centering
		\includegraphics[width=\textwidth]{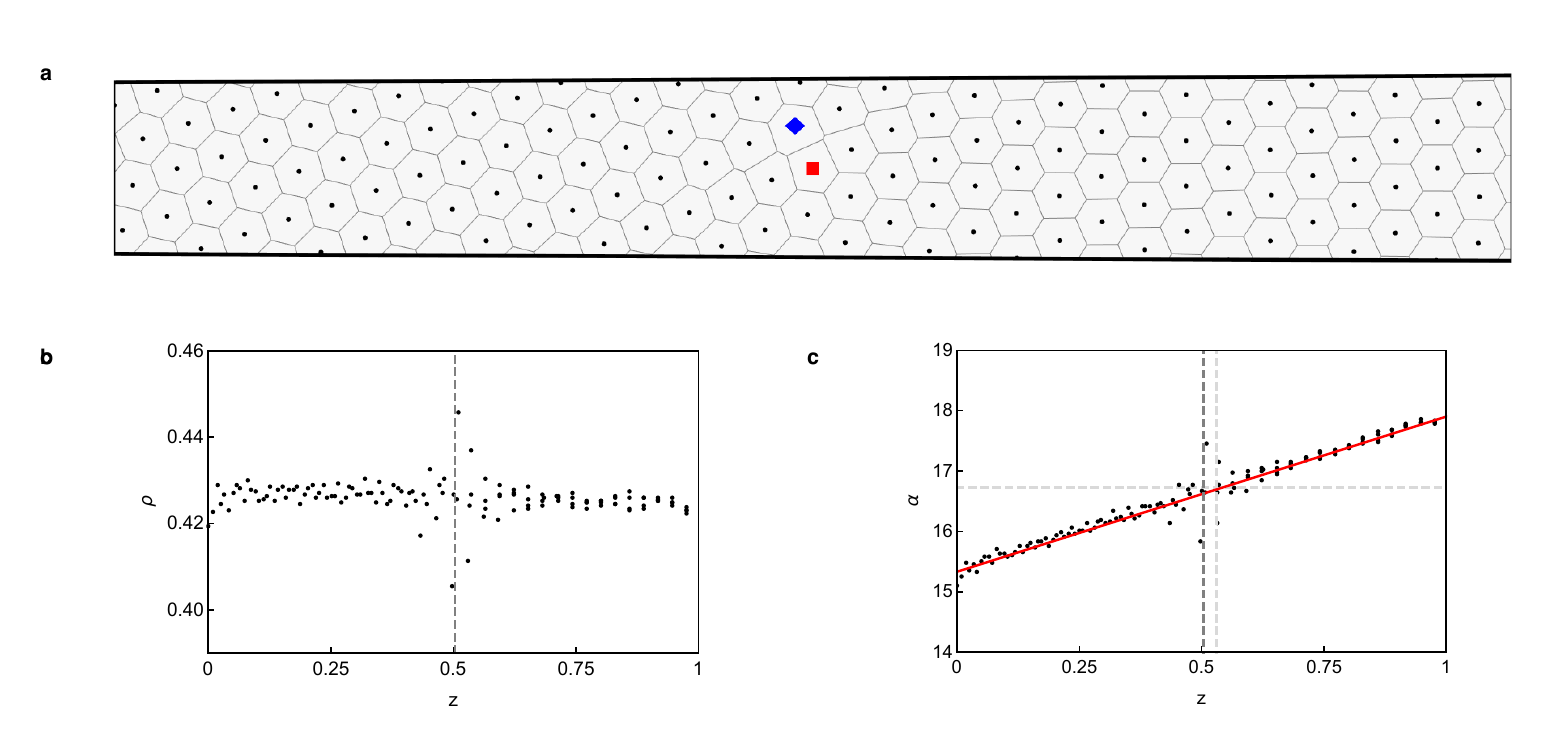}
		\caption{Ground state structure obtained using MD methods. (a) Snapshot of the system showing the \((4,3,1)\) to \((4,4,0)\) transition. (b) Local density variation throughout the system. The dashed grey line indicates the position of the transition. (c) Local \(\alpha\) variation throughout the system. The dark grey line indicates the position of the transition while the lighter lines indicate the expected \(\alpha\) value of the transition and its corresponding position within the system. }
	\end{figure}
	
	\begin{figure}[!htb]
		\centering
		\includegraphics[width=\textwidth]{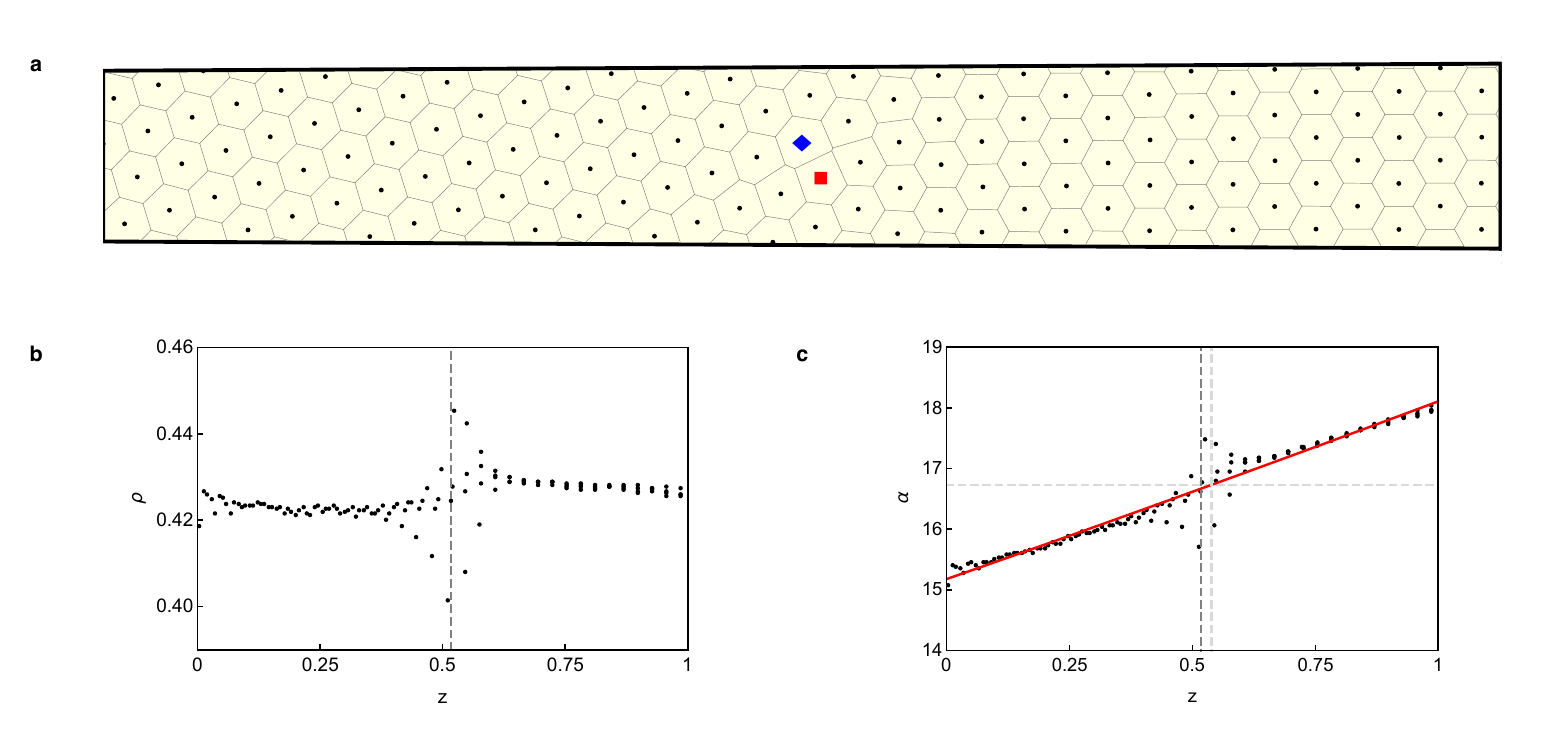}
		\caption{Ground state structure obtained using MC methods. (a) Snapshot of the system showing the \((4,3,1)\) to \((4,4,0)\) transition. (b) Local density variation throughout the system. The dashed grey line indicates the position of the transition. (c) Local \(\alpha\) variation throughout the system. The dark grey line indicates the position of the transition while the lighter lines indicate the expected \(\alpha\) value of the transition and its corresponding position within the system. }
	\end{figure}

\newpage
\newpage
\section{Molecular dynamics on the horn}
We parameterise the horn surface using cylindrical coordinates \((r, \theta, z)\), with positions on the surface described by \(\vb{r} = (r(z)\cos{\theta}, r(z)\sin{\theta},z)\). The surface has a profile
	\begin{equation}
		r(z) = \frac{r_0}{\sqrt{1-\eta z}}
		\label{eq:r}
	\end{equation}
where \(r_0 = c_0/(2\pi)\) is the radius at \(z=0\) and \(z\) is the fractional distance along the length of the surface, such that \(0<z<1\), and \(\eta>0\). 	

Throughout this section subscripts \(z\) or \(\theta\) indicate a derivative with respect to that variable, while subscripts \(i\) and \(j\) are used to index particular vortices. It is useful to note the following relations
	\begin{align}
		r_z(z) &= \frac{r_0\eta}{2(1-\eta z)^{3/2}} = \frac{\eta }{2r_0^2} r(z)^3\label{eq:rp}\\
		r_{zz}(z) &= \frac{3r_0\eta^2}{4(1-\eta z)^{5/2}} = \frac{3\eta^2 }{4r_0^4}r(z)^5. \label{eq:rpp}
	\end{align}

In order to do molecular dynamics, we need to know the geodesics on the surface between pairs of points. The distance \(d_{ij}\) between a pair of vortices is the length of the geodesic connecting the pair along the surface, a derivation for which is given in the following section. We numerically solve for the geodesics and calculate \(d_{ij}\) for each pair within the cut-off radius.  

The net force on a chosen vortex in a given time step is the sum of the individual forces acting upon it. We set a distance length scale \(a_0\) as the length of the lattice parameter for the expected state at \(z=0.5\), i.e. \(f_{ij} = -f_0 K_1(d_{ij}/a_0)\), where \(f_{ij}\) is the magnitude of the force on vortex \(i\) due to vortex \(j\), \(f_0=1\) is a constant and \(K_1\) is a modified Bessel function of the second kind. The distance \(\delta d\) moved by a vortex is restricted to a small value through the choice of the value of time step \(\delta t\). We therefore approximate that the vortex moves a distance \(\delta d\) along the tangent vector to the net force, which is approximately parallel to the geodesic along the direction of the net force over the distance \(\delta d\).

\subsection{Determining the distance between two particles - boundary value problem}
The distance between two points is determined by numerically calculating the length of the geodesic connecting the pair. Depending on the state on the surface, there may be occasions where either \(\theta_z\) or \(z_\theta\) is singular. As such, geodesic equations for both \(\theta(z)\) and \(z(\theta)\) must be known so that if one has a singularity, the other can be used.
\subsubsection{Solving for \(\theta(z)\)}
We determine the equation satisfied by the geodesic \(\theta(z)\) on the surface. The line element \(\dd s\) is given by
	\begin{align}
		\dd{s} &= \sqrt{\dd{r}^2+r^2\dd{\theta}^2+\dd{z}^2}\\
			&= \dd{z}\sqrt{r_z^2+r^2\theta_z^2+1}\\
			&\equiv \dd{z} f
	\end{align}	
where we have dropped the functional dependence \(r=r(z)\) for brevity.
		
The geodesic must satisfy the Euler-Lagrange equation for \(\theta\), leading to:
	\begin{align}
		0&=\dv{z}\pdv{f}{\theta_z} \\
		&= \frac{2rr_z\theta_z+r^2\theta_{zz}}{\sqrt{1+r_z^2+r^2\theta_z^2}} - \frac{r^2\theta_z(r_zr_{zz}+rr_z\theta_z^2+r^2\theta_z\theta_{zz})}{(\sqrt{1+r_z^2+r^2\theta_z^2})^3}
	\end{align}	
Using that \(\sqrt{1+r_z^2+r^2\theta_z^2}>0\), we can multiply out the denominator to give
	\begin{align}
		0 = &\, (2rr_z\theta_z+r^2\theta_{zz})(1+r_z^2+r^2\theta_z^2) 
			- r^2\theta_z(r_zr_{zz}+rr_z\theta_z^2+r^2\theta_z\theta_{zz})\\ 		
		0= &\,r\left[r(1+r_z^2)\theta_{zz}+r^2r_z\theta_z^3 
		+(2r_z(1+r_z^2)-rr_zr_{zz})\theta_z\right]
	\end{align}	
	
We make use of Eqs.~(\ref{eq:r})-(\ref{eq:rpp}) leading to 
	\begin{align}
		0&=  \left(1+ \frac{\eta^2 r^6}{4r_0^4}\right)\theta_{zz}+\frac{\eta r^4}{2r_0^2}\theta_z^3+r^2\left(\frac{\eta}{r_0^2}-\frac{\eta^3r^6}{8r_0^6}\right)\theta_z\label{eq:de}
	\end{align}
Substituting in for \(r(z)\) then leads to the equation satisfied by the geodesic \(\theta(z)\)
	\begin{align}		
		0=&\, 2(1-\eta z)(4(1-\eta z)^3+ \eta^2r_0^2)\theta_{zz}  
		+4\eta r_0^2(1-\eta z)^2\theta_{z}^3+ \eta(8(1-\eta z)^3-\eta^2r_0^2)\theta_z \label{eq:thetazbvp}
	\end{align}
Equation~(\ref{eq:thetazbvp}) can be solved numerically, with boundary conditions given by the positions of the two vortices, to find \(\theta(z)\) and \(\theta_z(z)\). The distance \(d_{ij}\) between vortices \(i\) and \(j\) is then the length of the geodesic and is determined by numerically integrating
	\begin{align}
		d_{ij} = \int \dd{s} = \int_{z_i}^{z_j} \dd{z}\sqrt{1+r_z(z)^2+r(z)^2\theta_z(z)^2}.
	\end{align}	


\subsubsection{Solving for \(z(\theta)\)}	
We repeat the process from the last section, this time solving for the geodesic \(z(\theta)\). In this case the line element \(\dd s\) is given by
	\begin{align}
		\dd s & = \dd{\theta}\sqrt{r(z(\theta))^2+z_\theta^2+r_\theta(z(\theta))^2}\\
			& = \dd{\theta}\sqrt{\frac{r_0^2}{1-\eta z}+z_{\theta}^2+\frac{r_0^2\eta^2 z_{\theta}^2}{4(1-\eta z)^3}}\\
			& \equiv \dd{\theta} g
	\end{align}
			
Solving the Euler-Lagrange equation for \(z\) gives
	\begin{align}
		&\pdv{g}{z} = \dv{\theta}\pdv{g}{z_{\theta}}\\
		&\pdv{g}{z} = \frac{\frac{3\eta^3r_0^2 z_{\theta}^2}{4(1-\eta z)^4}+\frac{\eta r_0^2}{(1-\eta z)^2}}{2\sqrt{\frac{\eta ^2r_0^2 z_{\theta}^2}{4(1-\eta z)^3}+\frac{r_0^2}{1-\eta z}+z_{\theta}^2}}\\	
		&\dv{\theta}\pdv{g}{z_{\theta}}  = 	\frac{\frac{\eta ^2 r_0^2 z_{\theta\theta}}{2 (1-\eta  z)^3}+\frac{3 \eta^3 r_0^2 z_{\theta}^2}{2 (1-\eta z)^4}+2 z_{\theta\theta}}{2 \sqrt{\frac{\eta^2 r_0^2 z_{\theta}^2}{4 (1-\eta  z)^3}+\frac{r_0^2}{1-\eta  z}+z_{\theta}^2}}
	-\frac{1}{4}\left(\frac{\eta ^2 r_0^2 z_{\theta}}{2 (1-\eta  z)^3}+2 z_{\theta}\right)\times\nonumber\\ &\left(\frac{3 \eta^3 r_0^2 z_{\theta}^3}{4 (1-\eta  z)^4} +\frac{\eta r_0^2 z_{\theta}}{(1-\eta  z)^2}\right. +\left.\frac{\eta ^2 r_0^2 z_{\theta} z_{\theta\theta}}{2 (1-\eta  z)^3}+2 z_{\theta} z_{\theta\theta}\right)
	\left(\frac{\eta ^2 r_0^2 z_{\theta}^2}{4 (1-\eta  z)^3}+\frac{r_0^2}{1-\eta  z}+z_{\theta}^2\right)^{-3/2}	
	\end{align}	
Simplifying this leads to the equation for the geodesic for \(z(\theta)\):
	\begin{align}
		0 = &\,2 (1-\eta  z) \left(4 (1-\eta  z)^3+\eta ^2 r_0^2\right)z_{\theta\theta} 
		+\eta\left(\eta^2 r_0^2-8 (1-\eta  z)^3\right)z_{\theta}^2-4 \eta r_0^2 (1-\eta z)^2
		\label{eq:z(theta)}
   	\end{align}
		
Equation~(\ref{eq:z(theta)}) can be solved numerically to find \(z(\theta)\) and \(z_{\theta}(\theta)\). The distance \(d_{ij}\) between a vortex pair is once again the length of the geodesic and is determined by numerically integrating
	\begin{align}
		d_{ij} = \int \dd{s} = \int_{\theta_i}^{\theta_j} \dd{\theta}\sqrt{r(z(\theta))^2+z_\theta(\theta)^2+r_\theta(z(\theta))^2}
	\end{align}


\section{Determining the expected vortex locations on the horn}
\renewcommand{\theequation}{\thesection.\arabic{equation}}
We derive the equation of the set of curves which can be used to construct the phyllotactic state with constant \(\alpha\) on the curved surfaces. The vertices at which these curves intersect define the expected locations of vortex sites. The curves, known as \textit{loxodromes}, always have their tangent vector at a fixed angle relative to the parallels and meridians of the surface. For the surface of revolution of the curve \(r(z)\) about the \(z\) axis, parameterised in cylindrical coordinates as \(\vb{r} = (r(z)\cos{\theta}, r(z)\sin{\theta},z)\), a set of orthonormal unit vectors along the surface can be defined as
	\begin{equation}
		\vu{e}_{\theta} = \mqty(-\sin{\theta}\\\cos{\theta}\\0) \qquad \vu{e}_v = \frac{1}{\sqrt{1+r_z^2}}\mqty(r_z\cos{\theta}\\r_z\sin{\theta}\\ 1)
	\end{equation}	
where, in this section, subscript \(z\) denotes a derivative with respect to \(z\). The parallels of the surface are the lines of constant \(v\) (or equivalently \(z\)) and the meridians are the lines of constant \(\theta\).	

A curve \(\gamma(z)\) for which the tangent vector is always at some fixed angle \(\beta\) relative to the meridians must always be parallel to the curve
	\begin{equation}
		\vu{\beta} = \cos{\beta}\vu{e}_v+\sin{\beta}\vu{e}_{\theta}.
	\end{equation}
	The values of \(\beta\) used to construct a state are determined by the chosen value of \(\alpha\) and the corresponding lattice vector directions. 
	
Solving \(\vu{T} = \vu{\beta}\), where \(\vu{T}\) is the unit tangent vector to \(\gamma(z)\) means that \(\gamma(z)\) must satisfy
	\begin{align}
		\frac{1}{\sqrt{1+r_z^2+r^2\gamma^{2}_z}} \mqty(r_z\cos{\gamma}-r\gamma_z\sin{\gamma} \\ r_z\sin{\gamma}+r\gamma_z\cos{\gamma}\\ 1)
		= \frac{\cos{\beta}}{\sqrt{1+r_z^2}}\mqty(r_z\cos{\gamma}\\r_z\sin{\gamma}\\1)
+ \sin{\beta}\mqty(-\sin{\gamma}\\\cos{\gamma}\\0) 	\end{align}
where \(r(z)\) and \(r_z(z)\) are defined in Eqs.~(\ref{eq:r}) and (\ref{eq:rp}) respectively.
		
Solving for \(\gamma_z(z)\) leads to 
	\begin{equation}
		\gamma_z(z) = \pm\tan{\beta}\frac{\sqrt{1+r_z(z)^2}}{r(z)}.
	\end{equation}
This expression is true for the loxodromes on any surface of revolution with a shape profile \(r(z)\).  We integrate this expression on the horn with the boundary condition that \(\gamma(z=0)=0\) to give the solution
	\begin{align}
		\gamma(z) &=\frac{1}{3} \tan{\beta} \left[\sqrt{1+\frac{4}{\eta ^2 r_0^2}}-\sqrt{1+\frac{4 (1-\eta  z)^3}{\eta ^2
   r_0^2}}
   -\tanh
   ^{-1}\left(\sqrt{1+\frac{4}{\eta ^2 r_0^2}}\right)+\tanh ^{-1}\left(\sqrt{1+\frac{4 (1-\eta  z)^3}{\eta ^2 r_0^2}}\right)\right].
	\end{align}